# Calculating the Energy Band Structure Using Sampling and Green's Function Techniques

Milad Khoshnegar, Amir Hossein Hosseinia, Nima Arjmandi, and Sina Khorasani

*School of Electrical Engineering, Sharif University of Technology,*
*P. O. Box 11365-9363, Tehran, Iran*

*ABSTRACT* − In this paper, a new method based on Green's function theory and Fourier transform analysis has been proposed for calculating band structure with high accuracy and low processing time. This method utilizes *sampling* of potential energy in some points of crystal's unit cell with Dirac delta functions, then through lattice Fourier transform gives us a simple and applicable formula for most of nanostructures. Sampling of potential in a crystal lattice of any kind contains accurate approximation of actual potential energy of atoms in the crystal. The step forward regarding the method concentrated on two novel ideas: Firstly, the potential was sampled and approximated by delta functions spread over the unit cell. Secondly, the principal equation of lattice is translated into reciprocal lattice and resulted in a huge reduction of calculations. By this method, it is possible to extract the band structure of any one-, two- or three-dimensional crystalline structure.

## I. INTRODUCTION

**B**AND structure of a material is very important to calculating the materials property, specially for obtaining the electrical properties. Usually we need to have some information about the bands to simulate electron devices.

Unfortunately there is not any analytical solution for the Schrödinger's equation in a crystal. Even it is not usually an easy and straightforward task to calculate it using existing numerical schemes because of the huge complexity involved. Thus over the past years many approximation methods for calculating the band structure have been developed; these include the free electron approximation [5], nearly-free electron approximation [6], cellular method [7], [8], tight-binding method [9], augmented plane wave method [10], [11], Green's function method [12], [13], [14], scattering matrix method, orhtogonalized plane wave method [15], the Born Oppenheimer approximation [16], the Hartree- Fock-Slater approximation [17], and finally the so-called empirical Pseudo-potential method [18].

Properties of these methods and their description is not within our scope and the reader is refered to many text books and papers about them but briefly we can say that the free-electron methods are useful approximations to making insight to behavior of metals. Tight-binding is a very useful method for many materials such as covalent crystals, nanostructures, and empirical pseudopotential method is relatively accurate method for semiconductors if we know some of their properties by experiments. But they have a deficiency : They are both relatively complicated algorithms to approximate band structures which need a massive computation. With nowadays needs for photonic, acoustic and semiconductor band gap engineering and new materials, there is a necessity for calculating the band structure more easily and accurately. Specially there should be simple methods to make insight in to new structures such 1D Carbon nanotubes and 0D Carbon nanotori. Here we have extracted new methods instead of existing algorithms for approximating the band structure, which are applicable for all the periodic nanostructures and have a simple closed form together with simple mathematics that is easily computable. For bands of nanostructures at first we approximate the closed shell potential well by sampling through three-dimensional impulses and solve the Schrödinger's equation using the Green functions and applying Bloch condition for wave function. Then we simplify the equation of packet using some developed mathematical theorems for Dirac delta function and finally applying the condition for non-zero solution gives the dispersion equation.

## II. DEFINITIONS

We define transfer vectors in lattice domain $\boldsymbol{\alpha}$ and $\boldsymbol{\beta}$ as:

$$\boldsymbol{\alpha} = m\mathbf{a}_1 + n\mathbf{a}_2 + p\mathbf{a}_3 \qquad (A.1)$$
$$\boldsymbol{\beta} = m_1\mathbf{a}_1 + n_1\mathbf{a}_2 + p_1\mathbf{a}_3 \qquad (A.2)$$

which m, n, p and also m1, n1, p1 are integers, and $\mathbf{a}_i$ for ( i=1,2,3) is defined as the basic vector of lattice's unit cells.

In the same way we may define $\mathbf{G}_\alpha$ , $\mathbf{G}_\beta$ as transferring vectors in reciprocal lattice domain as:

$$\mathbf{G}_\alpha = 2\pi(m\mathbf{b}_1 + n\mathbf{b}_2 + p\mathbf{b}_3) \quad (A.3)$$

$$\mathbf{G}_\beta = 2\pi(m_1\mathbf{b}_1 + n_1\mathbf{b}_2 + p_1\mathbf{b}_3) \quad (A.4)$$

in which $\mathbf{b}_i$ ( i=1, 2, 3) defines reciprocal lattice's basic vectors. Also we need to identify the sampling points in lattice domain with a particular vector which contains the distance vector from definite atom in each unit cell:

$$\mathbf{r}'_{ns} = \mathbf{R}_{i\alpha} + \mathbf{R}_{ns} = \boldsymbol{\alpha} + \mathbf{S}_i + \mathbf{R}_{ns} \quad (A.5)$$

where $\mathbf{S}_i$ is the vector which shows location point of each atom in unit cell and $\mathbf{R}_{ns}$ is the distance vector of sampling point from $i$th atom.

Also we use Fourier transform vector in reciprocal lattice domain, which is defined in 3D form as:

$$\mathbf{h} = h_1\mathbf{b}_1 + h_2\mathbf{b}_2 + h_3\mathbf{b}_3 \quad (A.6)$$

### III. BLOCH'S THEORY & GREEN'S FUNCTION

We know from Bloch's theory that wave function of electron in a periodic lattice is the product of a linear phase term by a periodic wave function. For the periodic wave function we have:

$$\begin{aligned}\Phi_{\boldsymbol{\kappa}}(\mathbf{r}'_{ns}) &= \Phi_{\boldsymbol{\kappa}}(\mathbf{R}_{i\alpha} + \mathbf{R}_{ns}) \\ &= \Phi_{\boldsymbol{\kappa}}(\boldsymbol{\alpha} + \mathbf{S}_i + \mathbf{R}_{ns}) = \Phi_{\boldsymbol{\kappa}}(\mathbf{S}_i + \mathbf{R}_{ns})\end{aligned} \quad (B.1)$$

With referring to the Schrödinger equation in Rydberg units we get:

$$(\nabla^2 + \zeta)\Psi_{\boldsymbol{\kappa}}(\mathbf{r}) = V(\mathbf{r})\Psi_{\boldsymbol{\kappa}}(\mathbf{r}) \quad (B.2)$$

Now by application of Bloch's theorem

$$\Psi_{\boldsymbol{\kappa}}(\mathbf{r}) = \exp(-j\boldsymbol{\kappa}.\mathbf{r})\phi_{\boldsymbol{\kappa}}(\mathbf{r}) \quad (B.3)$$

to the equation (B.2), we get Schrödinger equation in the reduced form of:

$$((\nabla - j\boldsymbol{\kappa})^2 + \zeta)\phi_{\boldsymbol{\kappa}}(\mathbf{r}) = V(\mathbf{r})\phi_{\boldsymbol{\kappa}}(\mathbf{r}) \quad (B.4)$$

For solving above equation through Green's function we should note that sampling approximation of lattice's potential energy can be presented as below:

$$V(\mathbf{r}) = \sum_i \sum_{ns} c_{ns} \delta^{(3)}(\mathbf{r} - \mathbf{r}') \quad (B.5)$$

In fact, above relation is the sampled potential in a unit cell. Sigma with index $i$ counts atoms in an identified unit cell and $ns$ is the number of samples around each atom. $c_{ns}$ is the magnitude of potential energy in the sampling point.

From equation (A.10) we can find periodic wave function as:

$$\phi_{\boldsymbol{\kappa}}(\mathbf{r}) = -G_{\boldsymbol{\kappa}}(\mathbf{r}, \mathbf{r}_0) \otimes (V(\mathbf{r})\phi_{\boldsymbol{\kappa}}(\mathbf{r})) \quad (B.6)$$

where $\otimes$ is the convolution operator. Green's function satisfies:

$$[(\nabla - j\boldsymbol{\kappa})^2 + \xi]G_{\boldsymbol{\kappa}}(\mathbf{r}, \mathbf{r}_0) = -\delta^{(3)}(\mathbf{r} - \mathbf{r}_0) \quad (B.7)$$

Actually that the impulse response of a system with the Hamiltonian $\hat{H} = ((\nabla - j\boldsymbol{\kappa})^2 + \zeta)$ is $G_{\boldsymbol{\kappa}}(\mathbf{r}, \mathbf{r}_0)$. On the other hand equation (B.4) shows that the response of the system to $V(\mathbf{r})\phi_{\boldsymbol{\kappa}}(\mathbf{r})$ is $\phi_{\boldsymbol{\kappa}}(\mathbf{r})$, so by expanding (B.6):

$$\begin{aligned}\Phi_{\boldsymbol{\kappa}}(\mathbf{r}) &= G_{\boldsymbol{\kappa}}(\mathbf{r}, \mathbf{r}_0) \\ &\otimes [\sum_{\boldsymbol{\alpha}} \sum_i \sum_{ns} c_{ns} \delta^{(3)}(\mathbf{r} - \mathbf{r}'_{ns})\Phi_{\boldsymbol{\kappa}}(\mathbf{r})]\end{aligned} \quad (B.8)$$

results. Substituting series expansion of $\phi_{\boldsymbol{\kappa}}(\mathbf{r})$ gives

$$\begin{aligned}\sum_\alpha \phi_\alpha \exp(j\mathbf{G}_\alpha.\mathbf{r}) &= -G_{\boldsymbol{\kappa}}(\mathbf{r}, \mathbf{r}_0) \\ &\otimes \{\sum_{\alpha'} \sum_i \sum_{ns} c_{ns} \delta^{(3)}(\mathbf{r} - \mathbf{R}_{i\alpha'} - \mathbf{R}_{ns}) \\ &\times \phi_{\boldsymbol{\kappa}}(\mathbf{s}_i + \mathbf{R}_{ns})\}\end{aligned} \quad (B.9)$$

Due to periodicity of $\phi_{\boldsymbol{\kappa}}(\mathbf{r})$ its Fourier transform in reciprocal domain can be calculated using completeness equation.

### IV. COMPLETENESS EQUATION

**THEOREM 1.** Let $(\mathbf{b}_1, \mathbf{b}_2, \mathbf{b}_3)$ be the reciprocal space vectors of the structure, and $(\mathbf{a}_1, \mathbf{a}_2, \mathbf{a}_3)$ be the primitive vectors structure space, and $\boldsymbol{\alpha}' = m\mathbf{a}_1 + n\mathbf{a}_2 + p\mathbf{a}_3$. Then we can write:

$$\begin{aligned}\sum_{\alpha'} \exp(j\,\mathbf{h}.\boldsymbol{\alpha}') &= \\ (2\pi)^3 \{\sum_{\alpha'} \delta(h_1 + \mathbf{G}_{\alpha'}.\mathbf{a}_2) \\ \times \delta(h_2 + \mathbf{G}_{\alpha'}.\mathbf{a}_2)\delta(h_2 + \mathbf{G}_{\alpha'}.\mathbf{a}_2)\}\end{aligned} \quad (C.1)$$

where $\mathbf{h}$ and $\mathbf{G}_{\alpha'}$ were defined before.

**Proof.** It's obvious that $\mathbf{b_j}.\mathbf{a_i} = 0$ if $i \neq j$ and $h_i = \mathbf{h}.\mathbf{a_i}$. So we can write:

$$\sum_{\alpha'} \exp(j\,\mathbf{h}.\boldsymbol{\alpha}') = (\sum_{m_1} \exp(jm_1 h_1))$$
$$\times (\sum_{n_1} \exp(jn_1 h_2))(\sum_{p_1} \exp(jp_1 h_3)) \quad \text{(C.2)}$$

Considering an infinite range for $m_1$, $n_1$, $p_1$ - i.e. an enough big crystal with dimensions bigger than approximately ten lattice constants or structures such as a nanotube or nanotorus – and applying Fourier series theorem it's easy to show that:

$$\sum_{\alpha'} \exp(j\,\mathbf{h}.\boldsymbol{\alpha}') = (2\pi \sum_{m_1} \delta(h_1 - 2\pi m_1))$$
$$\times (2\pi \sum_{n_1} \delta(h_2 + 2\pi n_1))(2\pi \sum_{p_1} \delta(h_3 + 2\pi p_1))$$
$$= (2\pi)^3 \sum_{\alpha'} \delta(h_1 + \mathbf{G_{\alpha'}}.\mathbf{a_1}) \delta(h_2 + \mathbf{G_{\alpha'}}.\mathbf{a_2}) \quad \text{(C.3)}$$
$$\times \delta(h_3 + \mathbf{G_{\alpha'}}.\mathbf{a_3})$$

## V. EXTRACTING FINAL EQUATION

Considering Theorem 1, the Fourier transform of $\phi_\kappa(\mathbf{r})$ is in the form of:

$$F\{\phi_\kappa(\mathbf{r})\} = (2\pi)^3 \{\sum_\alpha \phi_\alpha \, \delta(h_1 - \mathbf{G_\alpha}.\mathbf{a_1})$$
$$\times \delta(h_2 - \mathbf{G_\alpha}.\mathbf{a_2})\delta(h_2 - \mathbf{G_\alpha}.\mathbf{a_3})\} \quad \text{(D.1)}$$

Now exploiting this transform to equation (B.9) we have:

$$(2\pi)^3 \{\sum_\alpha \phi_\alpha \, \delta(h_1 - \mathbf{G_\alpha}.\mathbf{a_1})$$
$$\times \delta(h_2 - \mathbf{G_\alpha}.\mathbf{a_2})\delta(h_2 - \mathbf{G_\alpha}.\mathbf{a_3})\}$$
$$= -G_\kappa(\mathbf{h},\mathbf{r_0}).\{\sum_{\alpha'} \sum_i \sum_{ns} c_{ns} \, \phi_\kappa(\mathbf{s_i} + \mathbf{R_{ns}}) \quad \text{(D.2)}$$
$$\times \exp(-j\,\mathbf{h}.(\boldsymbol{\alpha}' + \mathbf{s_i} + \mathbf{R_{ns}}))\}$$

in which $G_\kappa(\mathbf{h},\mathbf{r_0})$ is the Fourier transform of Green's function. Referring to (B.7) we get:

$$G_\kappa(\mathbf{h},\mathbf{r_0}) = \frac{-\exp(-j\,\mathbf{h}.\mathbf{r_0})}{\zeta - |\mathbf{h}-\boldsymbol{\kappa}|^2} \quad \text{(D.3)}$$

Notice that in Rydberg units $K^2 = \zeta$. Also, given that $\zeta < 0$ we have $K = -jk$.

In equation (D.2) the exponential term on the right side can be divided into two terms that one has no dependence on $\boldsymbol{\alpha}'$. Finally exposing completeness equation to the other exponential term we have:

$$(2\pi)^3 \{\sum_\alpha \phi_\alpha \, \delta(h_1 - \mathbf{G_\alpha}.\mathbf{a_1})$$
$$\delta(h_2 - \mathbf{G_\alpha}.\mathbf{a_2})\delta(h_2 - \mathbf{G_\alpha}.\mathbf{a_3}) =$$
$$- G_\kappa(\mathbf{h},\mathbf{r_0}).(2\pi)^3 \{\sum_{\alpha'} \sum_i \sum_{ns} c_{ns} \, \phi_\kappa(\mathbf{s_i} + \mathbf{R_{ns}})$$
$$\times \exp(-j\,\mathbf{h}.(\mathbf{s_i} + \mathbf{R_{ns}})) \times \delta(h_1 + \mathbf{G_{\alpha'}}.\mathbf{a_2})$$
$$\times \delta(h_2 + \mathbf{G_{\alpha'}}.\mathbf{a_2})\delta(h_2 + \mathbf{G_{\alpha'}}.\mathbf{a_3})\}$$
$$\quad \text{(D.4)}$$

As we have some delta functions on both sides of (D.4), if $\boldsymbol{\alpha} = -\boldsymbol{\alpha}'$ then equating their weight we obtain:

$$\phi_\alpha = -\{\sum_i \sum_{ns} c_{ns} \, G_\kappa(\mathbf{G_\alpha},\mathbf{r_0}) \, \phi_\kappa(\mathbf{s_i} + \mathbf{R_{ns}})$$
$$\exp(-j\,\mathbf{G_\alpha}.(\mathbf{s_i} + \mathbf{R_{ns}}))\} \quad \text{(D.5)}$$

On the other hand from (D.3) we can find:

$$G_\kappa(\mathbf{G_\alpha},\mathbf{r_0}) = \frac{\exp(-j\mathbf{G_\alpha}.\mathbf{r_0})}{k^2 + |\mathbf{G_\alpha} - \boldsymbol{\kappa}|^2} \quad \text{(D.6)}$$

Now our final equation can be derived if we expand $\phi_\kappa(\mathbf{s_i} + \mathbf{R_{ns}})$ as a Fourier series:

$$\phi_\alpha = -\frac{1}{k^2 + |\mathbf{G_\alpha} - \boldsymbol{\kappa}|^2}$$
$$\{\sum_\beta \sum_i \sum_{ns} c_{ns} \exp(-j\mathbf{G_\alpha}.\mathbf{r_0}) \quad \text{(D.7)}$$
$$\times \exp(-j\mathbf{G_\alpha}.(\mathbf{s_i} + \mathbf{R_{ns}}))$$
$$\times \phi_\beta \exp(j\mathbf{G_\beta}.(\mathbf{s_i} + \mathbf{R_{ns}}))\}$$

This however can be recast into the form

$$\phi_{mnp} = -\frac{1}{k^2 + |\mathbf{G_{mnp}} - \boldsymbol{\kappa}|^2}$$
$$\times (\sum_{m_1} \sum_{n_1} \sum_{p_1} \sum_i \sum_{ns} c_{ns} \, \phi_{m_1 n_1 p_1} \exp(-j\,\mathbf{G_{mnp}}\mathbf{r_0}) \quad \text{(D.8)}$$
$$\times \exp(j\,(\mathbf{G_{m_1 n_1 p_1}} - \mathbf{G_{mnp}}).(\mathbf{s_i} + \mathbf{R_{ns}})))$$

Based on the above coefficients, the periodic wave function can be written as:

$$\phi_{\kappa} = -(\sum_{\alpha}\sum_{i}\sum_{ns} \frac{c_{ns}}{k^2+|\mathbf{G}_\alpha - \boldsymbol{\kappa}|^2} \phi_{\kappa}(\mathbf{s}_i + \mathbf{R}_{ns}) \quad \text{(D.9)}$$
$$\times \exp(-j\,\mathbf{G}_\alpha.(\mathbf{s}_i + \mathbf{R}_{ns}))\exp(j\,\mathbf{G}_\alpha.(\mathbf{r}-\mathbf{r}_0)))$$

and so the Bloch's wave function takes the form:

$$\Psi_{\kappa}(\mathbf{r}) = -\{\sum_{\alpha}\sum_{i}\sum_{ns} \frac{c_{ns}}{k^2+|\mathbf{G}_\alpha - \boldsymbol{\kappa}|^2}$$
$$\times \phi_{\kappa}(\mathbf{s}_i + \mathbf{R}_{ns})\exp(-j\,\mathbf{G}_\alpha.(\mathbf{s}_i + \mathbf{R}_{ns} + \mathbf{r}_0)) \quad \text{(D.10)}$$
$$\times \exp(-j\,(\boldsymbol{\kappa} - \mathbf{G}_\alpha).\mathbf{r})\}$$

Actually we are seeking for eigenvalues of energy and we are not looking for the wave function among the structure, so in the next step our consideration is focused on extracting $\phi_\alpha$ coefficients. For this purpose it's better to rewrite (D.7) in the closed form of:

$$\zeta.\phi_\alpha = |\mathbf{G}_\alpha - \boldsymbol{\kappa}|^2 \phi_\alpha - \sum_{\beta} a_{\alpha,\beta}\,\phi_\beta \quad \text{(D.11)}$$

in which $a_{\alpha,\beta}$ are defined as:

$$a_{\alpha,\beta} = \{\sum_{i}\sum_{ns} c_{ns} \exp(-j\,\mathbf{G}_\alpha.\mathbf{r}_0) \quad \text{(D.12)}$$
$$\times \exp(j\,(\mathbf{G}_\beta - \mathbf{G}_\alpha).(\mathbf{s}_i + \mathbf{R}_{ns}))\}$$

## VI. SIMULATION RESULTS

For approximating $\phi_{\kappa}(\mathbf{r})$ it's obvious that expansion of our lattice in simulation process must be limited by finite transferring vectors. As $m$, $n$, $p$ and also $m_1$, $n_1$, $p_1$ identify our expansion length in all 3 dimensions, they must be truncated to appropriate integer numbers. This truncation number depends on calculation volume and needed accuracy and should be optimized due to our limitations.

In fact our simulation process contains finding eigenvalues of $[\Phi]$ matrix which can be estimated from:

$$\zeta.[\Phi] = (\mathbf{B}-\mathbf{A})[\Phi] \quad \text{(E.1)}$$

where **B** and **A** are extracted from equation (D.11).

The other case which is necessary to be discussed is our sampling method. In the present analysis we create a semi-actual lattice based on the actual structure of desired crystal, by considering potential energy of each atom and also the effect of neighbor atoms. Dependening on how much accuracy we need, more neighbors even with higher distances could be considered. It is recommended to approximate the potential in a Wigner-Seitz cell and noticing that it is applicable for all cells in 3D expansion. In the sampling process we assume a finite and definite charge density for each closed-shell and we show that with $Q_{eff}$.

As a matter of fact this charge should be calculated in desired structure and for each special material we are analyzing. But here we just focus on the performance of our formula under simulation and it is not important for us to know about realistic measures of delta function weights. So the vertical axis of both potential and energy curves should be considered as normalized.

Firstly suppose that we have a 1D periodic structure (such as Krönig-Penny's model). Sampling its potential in a period gives us a curve such like that shown in Figure 1.

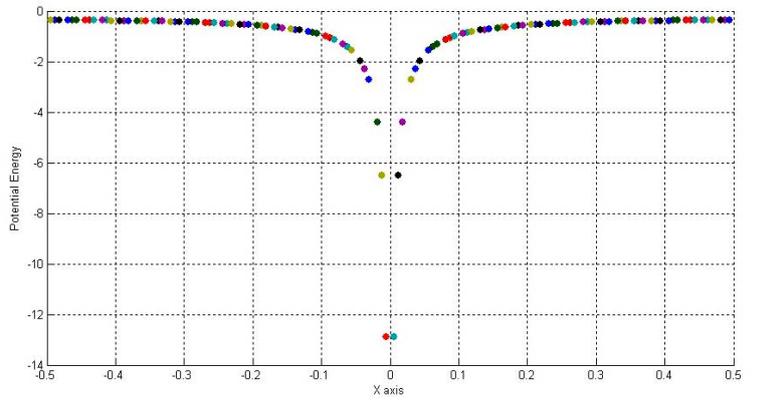

**Figure 1. Sampled potential around an identified atom in a 1D chain of atoms. The sampling process is performed in both right and left sides of desired atom to a length of half a lattice constant ($a$=1). X axis shows distance from atom which is located in X=0.**

For such a lattice explained above the band structure is shown in Figure 2. Three curves shown in this figure are the three first energy bands calculated by MATLAB program.

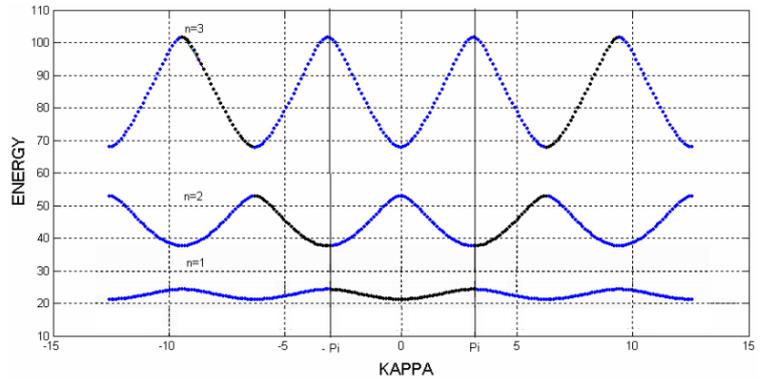

**Figure 2. Energy bands calculated for an atomic chain (Krönig-Penney model). Note that the lattice constant is normalized to unity ($a$=1).**

It is noteworthy to point out that within a denormalization scaling constant, this result is in complete accordance with our other new method [19], which is based on Wavelets.

For the 3D analysis we consider a simple cubic crystal. In this structure for each atom there are 6 first neighbors which have the most impact on potential weights of Delta functions at sampling points. As mentioned before we can consider even more neighbors (i.e second neighbors) for higher accuracy.

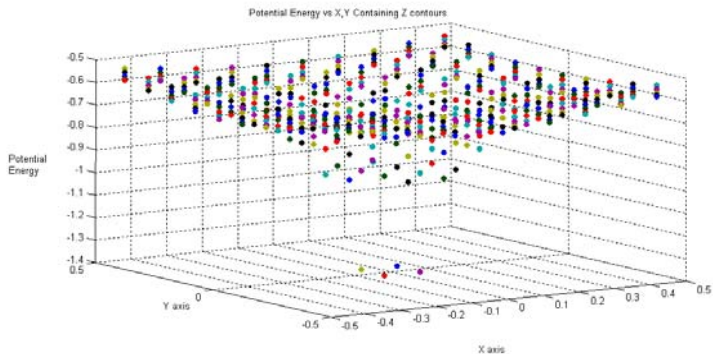

**Figure 3. Sampled potential energy for a simple cubic crystal, containing Z contours for each (X,Y) point. More negative potential points are closer to central atom in Wigner-Seitz cell.**

In Figure 3 we can see the sampled potential in a Wigner-Seitz cell of simple cubic structure. Colored dots in this figure show location of sampling points. Notice as we have 3D lattice potential energy is showed for contours of Z for each X and Y coordinate. Here, the sampling is uniform in space, however, for achieving higher accuracy one would need in general to take non-uniform samples from the potential in space. As it is obvious closer points to the central atom have more negative potentials.

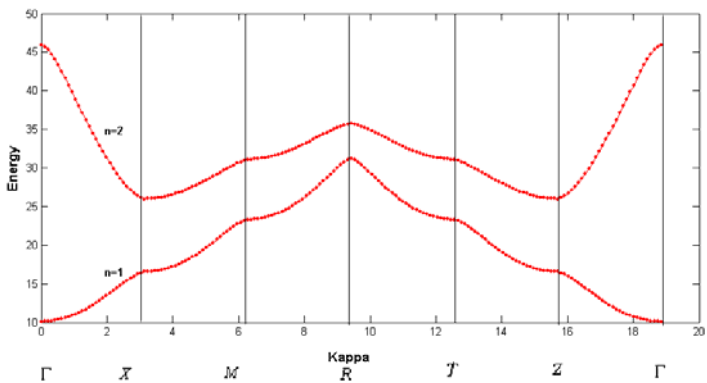

**Figure 4. First and second bands of a simple cubic lattice. Symmetry points, obvious in energy curves, are due to swept path shown in Figure 5.**

In Figure 4 we can see two energy bands for the band structure of simple cubic crystal. For creating this curve we swept the six part path shown in Figure 7. Notice to the symmetry points in our Brillion zone and also in our energy curves.

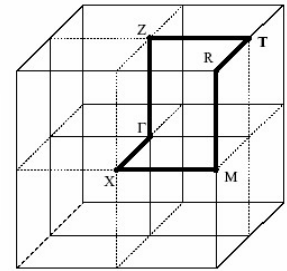

**Figure 5. Brillion zone for a simple cubic crystal and the path swept for calculating eigenvalues [1].**

## VII. CONCLUSION

We introduced a new, accurate and fast method for calculating band structure of small or large periodic structures based on sampling potential energy of crystal. We defined Fourier transform of lattice in such a useful way to solve Schrödinger's equation through Green's function and estimate dispersion equation. On the other hand correspondence of consequent simulation results shows that this method can be applied for fast approximation of electronic band structures.